# Lattice deformation at the sub-micron scale: X-ray nanobeam measurements of elastic strain in electron shuttling devices

C. Corley-Wiciak,[1] M. H. Zoellner,[1] I. Zaitsev,[1] K. Anand,[1] E. Zatterin,[2]

Y. Yamamoto,[1] A. A. Corley-Wiciak,[1] F. Reichmann,[1] W. Langheinrich,[3] L. R.

Schreiber,[4] C. L. Manganelli,[1] M. Virgilio,[5] C. Richter,[6] and G. Capellini[1,7]

[1]*IHP – Leibniz-Institut für innovative Mikroelektronik, Im Technologiepark 25,*
*D-15236 Frankfurt(Oder), Germany*

[2]*ESRF – European Synchrotron Radiation Facility, 71, avenue des Martyrs, CS 40220,*
*38043 Grenoble Cedex 9, France*

[3]*Infineon Technologies Dresden GmbH und Co.KG, Dresden, Germany*

[4]*JARA-FIT Institute for Quantum Information, Forschungszentrum Jülich and RWTH Aachen University, Germany*

[5]*Department of Physics Enrico Fermi, Università di Pisa, Pisa 56126, Italy*

[6]*IKZ – Leibniz -Institut für Kristallzüchtung, Max-Born-Straße 2, D-12489 Berlin, Germany*

[7]*Dipartimento di Scienze, Universita Roma Tre, Viale G. Marconi 446, Roma 00146, Italy*

(*Electronic mail: corley@ihp-microelectronics.com, carsten.richter@ikz-berlin.de)

The lattice strain induced by metallic electrodes can impair the functionality of advanced quantum devices operating with electron or hole spins. Here we investigate the deformation induced by CMOS-manufactured titanium nitride electrodes on the lattice of a buried, 10 nm-thick Si/SiGe Quantum Well by means of nanobeam Scanning X-ray Diffraction Microscopy. We were able to measure TiN electrode-induced local modulations of the strain tensor components in the range of $2 - 8 \times 10^{-4}$ with ~60 nm lateral resolution. We have evaluated that these strain fluctuations are reflected into local modulations of the potential of the conduction band minimum larger than 2 meV, which is close to the orbital energy of an electrostatic quantum dot. We observe that the sign of the strain modulations at a given depth of the quantum well layer depends on the lateral dimensions of the electrodes. Since our work explores the impact of device geometry on the strain-induced energy landscape, it enables further optimization of the design of scaled CMOS-processed quantum devices.

# I. INTRODUCTION

The characterization of mechanical strain at the nanometer scale is essential for improving the homogeneity and electronic performance in advanced semiconductor-based electronic devices. [1, 2] In particular, it is of paramount importance to assess the impact of the fabrication processes on the mechanical stress generated by electrodes made from metals or metallic compounds. [3] Indeed, an electrode exerts a stress onto the underlying and neighboring semiconductor layers, thus inducing local lattice deformations, described by the spatially dependent strain tensor $\varepsilon(x, y, z)$. In turn, the lattice strain affects electronic properties such as charge carrier mobility, [4] band edge potential, [5] and tunnel couplings. [2]

This is particularly relevant for modern semiconductor-based quantum devices, such as qubits, which are still exceedingly difficult to fabricate in large numbers and require a high degree of homogeneity in their material environment. [6] Spin qubits housed in electrostatic quantum dots (QDs) are under the spotlight since their fabrication can be integrated with microelectronics foundry environments, [7, 8] leveraging the high maturity level of the Si-based complementary metal oxide semiconductor (CMOS) processes. [9] For this purpose, commonly used metals for manufacturing electrodes in small-scale device processing, such as Al, Pd and Pt [10, 11, 12], should be replaced by materials commonly used in CMOS foundries, such as Titanium Nitride (TiN), [13] which has the advantage of being a thermally stable and chemically mostly inert material with low resistance and moreover acts as a diffusion barrier. [14, 15]

Quantum processors based on electrostatic QDs rely on gate electrodes, thus this choice of material is particularly relevant with regards to their spatial homogeneity. As a matter of fact, the realization of large arrays of densely packed QDs is required in architectures allowing for error correction schemes. [16] Each error corrected logical qubit would involve operating thousands of physical qubits. [17] Long coherence times have already been demonstrated for electron spins housed in epitaxial Si/SiGe heterostructures, [18] and simultaneous operation of multiple semiconductor spin qubits has recently been reported. [19] The operation of qubits in large arrays requires spin-coherent communication pathways among them, [20, 21] which can be realized e.g. by means of the quantum bus (QuBus) architecture. A QuBus device, as depicted in **Figure 1a**, shuttles electrodes in a conveyor mode scheme using a propagating sinusoidal potential pulse. This is generated by modulating the voltages on an array of "clavier" electrodes. [12, 22] The shuttling takes place either by adiabatic Landau Zener transitions across an array of tunnel-coupled QDs or by adiabatic motion of a quantum dot, as described in detail by Seidler *et al.* [12] As shown in **Figure 1b,** clavier gate arrays can also be employed to establish several QDs that may contain single spin qubits. [10]

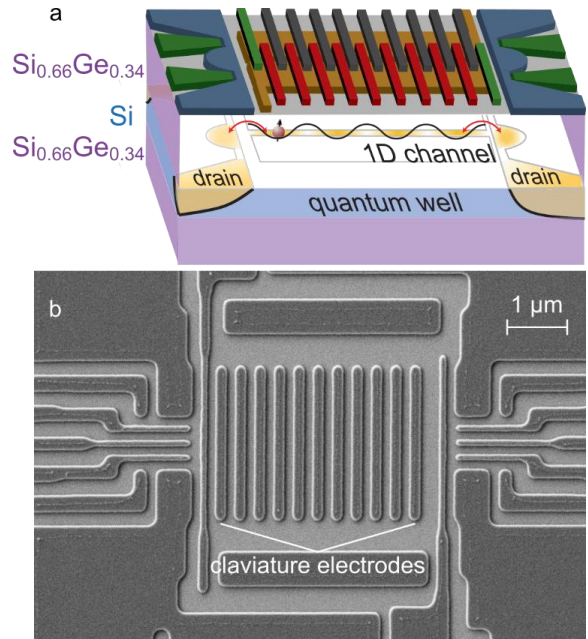

FIG. 1. **(a)** Schematic cross section of a QuBus device, reproduced with permission. *[22]* **(b)** Top-view SEM image around the clavier electrodes

The operation of several physical qubits and their interconnection by the QuBus will require a control scheme with voltages shared between the electrodes, [20] which does not allow for retuning of individual QDs. Therefore, the potential landscape in the active region of the device must feature a high degree of spatial homogeneity Thus, among other things, local mechanical deformations in the lattice caused by the electrodes should be avoided, due to their impact on the band edge potential. [23, 24, 6],

In general, the internal stress of a metal-compound thin film (or stripe) is comprised of thermomechanical and residual stress. [25] The thermomechanical stress is generated by the mismatch of the coefficient of thermal expansion (CTE) between the material and the semiconductor/oxide virtual substrate (VS) when the sample cools down from the deposition to the operation temperature. [2] The residual stress is accumulated during the deposition process owing to variations in crystallite sizes, orientations, relaxations, and coalescence. [25] In particular, stress exerted from TiN thin films on Si or SiGe/Si substrates may be tuned from tensile to compressive, [26, 27] by acting on layer thickness and deposition process parameters, [28, 3] This allows to tune the deposition process towards targeted stress values for specific applications.

The stress within large TiN films is typically investigated by wafer curvature measurements; [29] however, the stress in small lithographically fabricated electrodes may differ from the unstructured film due to amorphization, oxidation and plastic relaxation occurring during structuring. From this perspective, scanning measurements with focussed X-ray beams are well suited to study the local material properties in microelectronic devices. [30, 31]

Motivated by the above considerations, we have employed Scanning X-ray Diffraction Microscopy (SXDM) at the X-ray nanoprobe beamline ID01/ESRF, [32, 33] to obtain microscopic maps of the lattice strains induced by TiN electrodes deposited on the top surface of two samples with 10 nm-thick Si quantum wells (QW) grown on $Si_{0.66}Ge_{0.34}$ buffers on Si(001) substrates. More precisely, we compare the local strain distribution measured in two devices sharing the same epitaxial layer stack and device layout, but differing in the sputtering process used for TiN deposition, resulting in different global stress in the TiN layer.

To evaluate the impact of the strain fluctuations on the potential landscape of the QW, using the measured strain profiles we calculate the local perturbation of the Si conduction band edge in the framework of deformation potential theory. [34] Furthermore, we employ Finite Element Method (FEM) simulations with *COMSOL Multiphysics* to investigate the strain field geometry induced by electrodes of varying lateral dimensions in the underlying active layer.

## II. Experiment

The epitaxial layer stack of the samples, shown in **Figure 2a,** was grown on a commercial 200 mm Si wafer by Reduced Pressure Chemical Vapor Deposition (RP-CVD) and comprises of five 500 nm-thick step-graded $Si_{1-x}Ge_x$ layers with increasing Ge content $x$, a 4 μm-thick plastically relaxed $Si_{0.66}Ge_{0.34}$ buffer, a 10 nm pseudomorphic Si QW layer, a 30 nm $Si_{0.66}Ge_{0.34}$ barrier and a sacrificial Si cap (< 2 nm). The threading dislocation density (TDD) was measured to be $< 2 \cdot 10^6$ cm$^{-2}$ by pit counting after a Secco etch. After removing the cap layer for surface cleaning, a 10 nm $SiO_2$ dielectric layer was deposited on top of the semiconductor stack by CVD. Finally, a TiN film of approx. 30 nm thickness was deposited by sputtering PVD, followed by deposition of a $SiO_2$ hardmask used to fabricate the metal gate stripes by UV-lithography.

Two samples, A and B, are studied in this work. Sample A was sputtered with a Magnetron for a denser plasma, while Sample B was sputtered without the Magnetron. The average, global stress the TiN is subjected to, was determined by measuring the curvature of the samples with X-ray Diffraction (XRD) after processing and applying Stoney's equation (see the SM), [29] finding a stronger stress for sample A ($\sigma_A = -2.6$ GPa) and a weaker stress for sample B ($\sigma_B = -1.5$

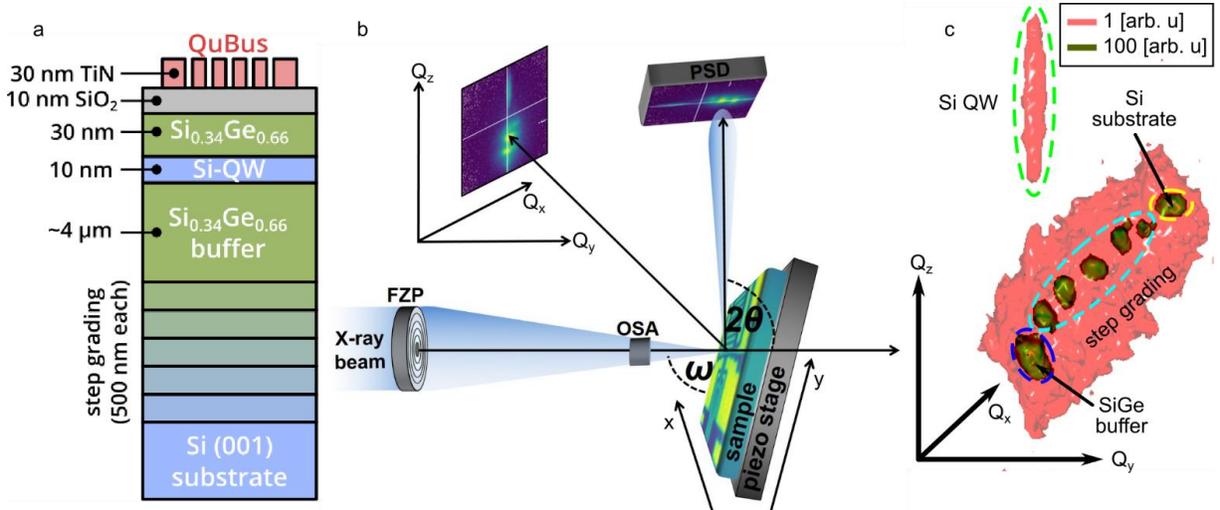

FIG. 2. **(a)** Heterostructure layer stack of the samples **(b)** Experimental setup of the SXDM measurement. **(c)** Isosurface plot of the 3D RSM around the 335 Bragg reflections of the heterostructure for one spot on the sample. The axes of reciprocal space are defined as $Q_x \parallel [1\bar{1}0]$ $Q_y \parallel [110]$, $Q_z \parallel [001]$.

GPa).

The Synchrotron measurements were carried out at ID01/ESRF with the setup depicted in **Figure 2b**. The X-ray beam was focused by a Fresnel Zone Plate (FZP) to a spot of ∼ 60 nm diameter on the sample surface, while the higher order diffraction from the FZP is blocked by an order sorting aperture (OSA). The spot size was verified by ptychographic reconstruction (see Supplemental Material (SM)). [35] The X-ray energy was set to 10 keV, at which the 335 reflection of Si is accessible in near-normal incidence and grazing exit of the X-ray beam. The high incidence angle of approx. 88° avoids any significant broadening of the beam footprint by projection on the sample surface, while the small exit angle of approx. 8° ensures a shallow penetration depth of the X-rays and thus high sensitivity to the QW layer close to the sample surface. Simultaneously to the diffraction, the fluorescence from the Ti K-edge is mapped with an energy-resolved X-ray detector, allowing a tracking of the TiN electrodes.

The experimental maps were acquired by scanning the sample in steps of ∼50 nm across the X-ray beam with a piezo stage, while the diffraction signal was recorded with a *Maxipix* area detector. Due to the improved brilliance after the Extremely Bright Source upgrade, [36] an exposure time of 20 ms per spot on the sample was sufficient to measure the 10 nm-thick QW layer diffraction signal with a signal-to-noise ratio > 10. To sample a three-dimensional (3D) volume of reciprocal space, diffraction maps are measured for a series of sample rocking angles $\omega$. In this way, from the diffraction patterns measured at every spot on the sample, maps of the local scattering vector have been generated (see the SM for details).

## III. Results

An exemplary 3D reciprocal space map (RSM) for one spot on the sample is shown in **Figure 2c.** The most intense peaks in the RSM originate from the Si substrate and the 4 µm-thick buffer layer with constant composition. Between them, five intensity peaks can be observed, stemming from the discrete composition steps of the graded buffer material. The signal Bragg peak corresponding to the Si QW is located above the $Si_{0.66}Ge_{0.34}$ signal, at equal in-plane momentum transfers $Q_x$, $Q_y$, meaning that the QW is pseudomorphic to the buffer layer. Remarkably, the QW signal can be clearly separated from the more intense signals of the buffer along $Q_z$ due to its smaller out-of-plane lattice constant.

We note that in our reference system, the orientation of the length constants of the unit cell, the planes of the direct and reciprocal lattice and coordinates of the maps is defined as:

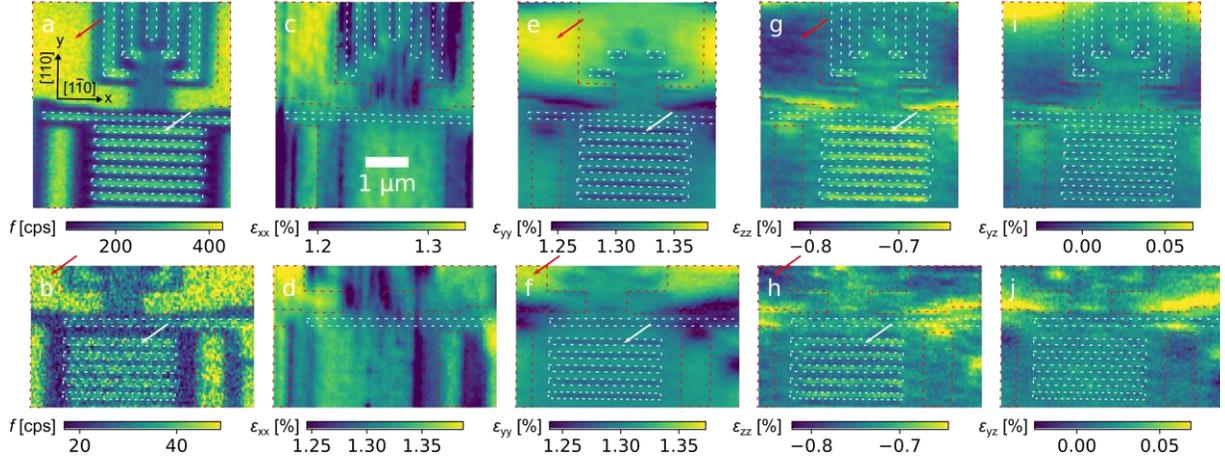

FIG. 3. SXDM maps of the the samples A (**top row**) and B (**bottom row**). The black arrows describe the orientation of the crystallographic axes, the scale is identical for all images. The clavier electrodes are located in the bottom region of the maps. Maps of (**a,b**) Ti K-edge fluorescence $f$ and strains in the Si QW layer: (**c,d**) $\varepsilon_{xx}$, (**e,f**) $\varepsilon_{yy}$, (**g,h**) $\varepsilon_{zz}$, (**i,j**) $\varepsilon_{yz}$. The device geometry is outlined by the dotted lines, with contact pads of large lateral width marked in red and narrow electrodes in white. Notice that orientation of the $(1\bar{1}0)$ and $(110)$ lattice planes is consistent with the observed direction of misfit dislocation (MD) bunches in strain maps for the relaxed SiGe buffer shown in the SM.

$a \parallel x \parallel Q_x \parallel [1\bar{1}0]$, $\quad b \parallel y \parallel Q_y \parallel [110]$,
$c \parallel z \parallel Q_z \parallel [001]$.

By combining diffraction maps of the $335$ and $\bar{3}35$ Bragg reflection, we can measure the lattice constants $b$, $c$, and the angle $\alpha$ of the Bravais cell in the Si QW, [37]. These data are used to generate maps in the direct space of the $\varepsilon_{yy}$, $\varepsilon_{zz}$ and $\varepsilon_{yz}$ components of the lattice strain tensor $\varepsilon$ relying on the following equations [38]

$$\varepsilon_{yy} = \frac{b \sin\alpha}{a_{Si}} - 1, \quad (1)$$

$$\varepsilon_{zz} = \frac{c}{a_{Si}} - 1, \quad (2)$$

$$\varepsilon_{yz} = \frac{1}{2}\frac{b \cos\alpha}{a_{Si}} \quad (3)$$

where $a_{Si}$ was set at the value of 5.4309 Å. [39] (the orientation of the coordinate system with respect to the cristallographic axes is specified in the first panel of Fig. 3 With a diffraction map of the $\bar{3}35$ reflection, we also determine the local value of the QW lattice constant $a$, and the corresponding strain $\varepsilon_{xx}$ in the small angle approximation as

$$\varepsilon_{xx} = \frac{a}{a_{Si}} - 1 \quad (4)$$

In **Figure 3**, maps of the Ti K-edge X-ray fluorescence $f$ and four components of the strain tensor are shown for the "high stress" sample A (left) and "low stress" sample B (right), respectively. The fluorescence is given in counts per second, the strain is dimensionless and given in percent. The maps cover slightly different areas of the QuBus on the two samples due to sample drift during the measurement (~1 μm total), which is caused by thermal settling and a residual offset between the beam focus and the $\omega$-rotation axis. This drift was tracked during the measurement by simultaneously observing the diffraction and fluorescence signals (**Fig. 3 a,b**).

From the symmetric in-plane strain components, we measure in the pseudomorphic Si QW a biaxial, global average strain corresponding to $\varepsilon_{xx} \approx \varepsilon_{yy} \approx 1.3$ %, as expected from the lattice mismatch between Si and the $Si_{0.66}Ge_{0.34}$ virtual substrate.

With respect to this average biaxial strain, in both datasets we observe a distribution of the lattice strain in the QW layer in the form of a "footprint" of the device, as evidenced by the dashed lines. By comparing the fluorescence maps for both samples to the strain maps, it is apparent that this strain modulation in the QW layer is due to the overlaying TiN device.

We point out that there is a pronounced difference between the effect on the strain in the QW brought either by wide pads (red arrows) or narrow electrodes (white arrows). Indeed, underneath the wide structures, there is a tensile modulation of the in-plane lattice strains $\varepsilon_{xx}$ (**Fig. 3 c,d**) and $\varepsilon_{yy}$ (**Fig. 3 e,f**), while under narrow TiN stripes, in particular the clavier electrodes outlined by the white dashed boxes, the in-plane strain modulation become compressive, mainly due to smaller values of the diagonal strain component which correspond to the direction orthogonal to the electrode axis. Thus, the lateral dimension of the lithographic structures appears to determine not just the magnitude, but also the quality of their effect on the lattice strain in the underlying layer.

In the QuBus, electrons are shuttled through the clavier electrodes. Thus, the lattice deformation underneath them is most important for the device and hence in the following we will focus on this region of the device.

In line with the above observations, the symmetric in-plane strain component $\varepsilon_{xx}$ (**Fig. 3c,d**) appears practically unaffected by the clavier electrodes which run along the $[1\bar{1}0]$ direction. However, across TiN stripes oriented along [110], we observe characteristic modulations in $\varepsilon_{xx}$ of several $10^{-4}$. Accordingly, the clavier electrodes have a compressive effect on $\varepsilon_{yy}$, with a total bandwidth of $\sim 4\times 10^{-4}$ for sample A (**Fig. 3e**) and $\sim 2\times 10^{-4}$ for B (**Fig. 3f**). This modulation is noticeably homogeneous both along each stripe and similar across different electrodes.

For the out-of-plane strain component, which has an average value of $\varepsilon_{zz} \approx$ -0.73 %, we also observe effects stemming from the TiN structures, but with reversed sign compared to the modulations of the in-plane strains: when the Si crystal lattice is strained along an in-plane direction by an electrode, it reacts in the vertical direction since it tends to preserve the volume.

Thus, we observe that $\varepsilon_{zz}$ becomes more positive underneath the clavier electrodes, with modulations of up to $\sim 8\times 10^{-4}$ for sample A (**Fig. 3g**) and up to $\sim 4\times 10^{-4}$ for B (**Fig. 3h**).

The average value of the off-diagonal strain component $\varepsilon_{yz}$ is close to zero, as expected for the case of a purely biaxial distortion, but there are local modulations connected to the truncation of the electrodes. Thus, the shape of the device is also well recognizable in the $\varepsilon_{yz}$ maps (**Fig. 3 i,j**). They show modulations across the clavier electrodes of a magnitude of approx. $\sim 2\times 10^{-4}$, clearly demonstrating a shear deformation of the lattice in the $yz$-plane.

We point out that strain maps obtained with the same technique here employed have been previously reported in Ref. [23], where a similar quantum structure has been investigated. However, in the device studied in Ref. [23], the quantum dots have been fabricated starting from a Ge/Si$_{0.2}$Ge$_{0.8}$ QW material, which required the adoption of a plastically relaxed Ge virtual substrate. As a side effect of this choice, the measured strain maps evidenced the presence of long range strain gradients, which were attributed to (1) the presence of the low-lying misfit dislocation networks in the VS formed to accommodate the lattice mismatch and (2) local fluctuations of Ge content in the SiGe/Si VS. [40] In this heterostructure, their magnitude is smaller due to the lower mismatch between the Si substrate and the Si$_{0.66}$Ge$_{0.34}$ buffer layer, as well as the large thickness of the VS in the samples studied here (> 5 μm total). This is in agreement with the findings of *Park et al.*, which found that the strain fluctuations caused by the MD-network formed during the plastic relaxation in a Si$_{0.7}$Ge$_{0.3}$/Si/Si$_{0.7}$Ge$_{0.3}$/Si heterostructure are smaller in comparison with the electrode-induced strains. [24, 31]

In fact, it is difficult to distinguish the effect of the MD network and the larger TiN pads around the narrow clavier gates within the comparatively small field of view of the SXDM measurement. Thus, we studied the effect of the MD network with μ-Raman shift mapping (see the SM). The μ-Raman data confirms that the magnitude of the corresponding modulation of the lattice strain is approx. $5\times10^{-4}$, but it takes place on a length scale larger than 1 μm, i.e. very long compared to the abrupt modulations induced by the clavier electrodes. We point also out that SXDM strain maps taken from the relaxed $Si_{0.66}Ge_{0.34}$ layers likewise confirm that the magnitude of the strain modulations due to the MD network is on the order of $\sim 10^{-4}$.

Leveraging on the relationship,

$$\nu = \frac{1}{1-\frac{\varepsilon_{xx}+\varepsilon_{yy}}{\varepsilon_{zz}}} \approx \frac{1}{1-\frac{C_{33}}{C_{13}}} \qquad (5)$$

which links the lattice strains, all measured by SXDM, to the Poisson number $\nu$ and the elastic coefficients $C_{13} = C_{23}$, $C_{33}$, [41, 42] we obtain an average value in the QW layer of $\nu_{QW} \approx 0.22$. This is significantly lower than the established value for bulk silicon ($\nu_{bulk} \approx 0.278$), [39, 41] and would correspond to a change in the elastic coefficients. Since a similar discrepancy has been observed also for epitaxial Ge films [42], one cause of this smaller Poisson number may be that the elastic properties and therefore in the thin, strained Si QW layer are different from those of bulk Silicon. Moreover, we point out that the QW is under a strong compressive biaxial stress with $\sigma_{xx} \approx \sigma_{yy} \approx -2.5$ GPa, as obtained from the measured strains using the elastic coefficients for Silicon reported in Ref. [41] while rotating the stiffness tensor to match the orientation of our reference system. [23] In this condition, given the QW layer thickness of ~10 nm, the Si material is at the critical point for the onset of plastic relaxation. [43] Therefore, non-linear elastic effects may become relevant, [44, 45] causing an apparent deviation of the Poisson number from its bulk value.

Furthermore, local volume compression or expansion triggered by the metal gates may occur within these samples. As a matter of fact, eq. 5 relies on the assumption of zero surface normal and shear stress ($\sigma_{zz} = \sigma_{xz} = \sigma_{yz} = 0$) a condition which may be not perfectly fulfilled in our case. In fact the stiffness of the electrodes, [23] and the local rotations of the lattice planes in the QW layer relative to the sample surface may lead a violation of this assumption. [46]

These effects may combine into some amount of local hydrostatic stress and volume change in the QW layer. As a consequence, we can observe the footprint of the QuBus device also in the maps of the local variations of stress along the surface normal $\sigma_{zz}$ shown in **Fig. 4**, obtained using Hooke's law with the elastic coefficients for Silicon in conjunction with the local lattice strains determined by the SXDM measurements: [47]

$$\sigma_{zz} = C_{13}\varepsilon_{xx} + C_{23}\varepsilon_{yy} + C_{33}\varepsilon_{zz} \qquad (6)$$

As for the strain maps, we observe local modulations induced by the TiN structures. In particular, the clavier electrodes appear to apply a tensile stress in the QW layer along the vertical

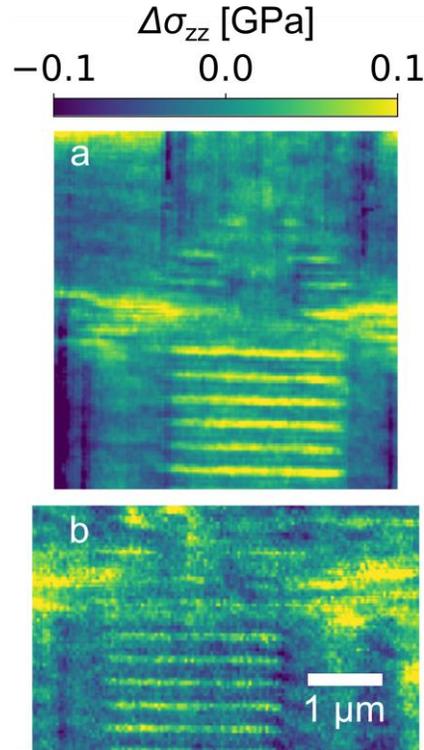

FIG. 4. Maps of the local variation of surface normal stress $\Delta\sigma_{zz}$, calculated from the strain maps for Hooke's law, for the samples A **(a)** and B **(b)**.

direction ($\sigma_{zz} > 0$).

In order to evaluate the impact of the electrode-driven lattice deformation on the energy landscape, we have determined in the framework of the deformation potential theory the strain-dependent conduction band minimum in the Si QW layer [34], since it represents a key ingredient for the calculation of the eigenvalue spectrum of few electrons QD systems [48]. Due to the average tensile biaxial strain in the Si layer, the bottom of the conduction band in this region is controlled by the strain-induced fluctuation of the $\Delta_2$ energy.

In **Figure 5**, line profiles taken across five clavier electrodes along the y direction are shown for the strain components $\varepsilon_{yy}$ (**panel a**) and $\varepsilon_{zz}$ (**panel b**), measured in the QW layer at room temperature. The amplitudes of the electrode-driven strain fluctuations are, on average, larger by a factor of approx. two for the high stress sample A compared to low stress sample B. This demonstrates that even after lithography and structuring, the deposition of a high stress TiN layer causes a stronger deformation in the underlying semiconductor compared to the TiN with weaker initial stress.

For what concerns the band edge, correspondingly we observe that the $\Delta_2$ energy (**panel c**) undergoes larger local modulations in sample A, for which we estimate a bandwidth of ±4 meV to be compared with the range of ±2 meV calculated for B. The maxima of the $\Delta_2$ level are located directly underneath the electrodes, while the minima are in the gap between them. Thus, without biasing the system potential barriers of up to approx. 8 meV for sample A and approx. 4 meV for B are already present between the electrodes.

Tunnel-coupled QDs formed by periodic metal gates can be optimized to achieve low variance $\Delta_{orb}$ of the orbital energy $E_{orb}$ (e.g. $E_{orb} \approx 3$ meV, $\Delta_{orb} \approx 0.2$ meV for an array of 9 QDs in Ref. [10]), suggesting a low impact of the strain potential. However, we suggest that several effects could increase the impact of the electrodes.

Firstly, the strain modulations we observed for the TiN electrodes are much stronger compared to those we previously measured in a device with metal (Ti/Pd, Al) electrodes. [23] This is due to the stronger stress (few GPa) in the TiN, compared to e.g. Pd electrodes, which typically experience smaller stresses (< 1 GPa). [49] Thus, the impact of the TiN electrode studied here on the potential level of the band edges is larger compared to gate electrodes fabricated from metals such as Pd and Pt. Second, the impact may be lowered by larger coverage, as usually at least two interleaved layers are used in quantum devices, which may lead to a more homogenous strain distribution in the active layer.

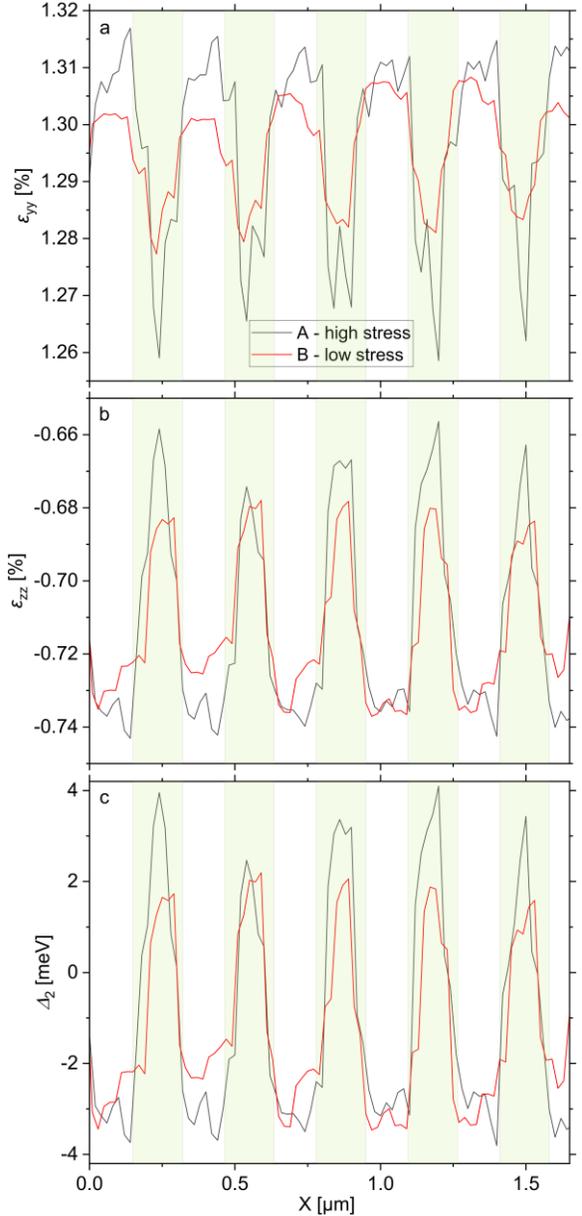

FIG. 5. Profiles across five clavier gate electrodes of **(a)** in-plane strain $\varepsilon_{yy}$, **(b)** out-of-plane strain $\varepsilon_{zz}$, and **(c)** conduction band energy minimum $\Delta_2$. The extent of the electrodes is shaded in green.

We also highlight that the spatial periodicity of the strain follows the periodicity of the gates and thus the one of the QDs. However, in a QuBus device operated in the conveyor-mode, a QD adiabatically propagating has to be formed at all positions along the QuBus. If gate biases are engineered without accounting for the unintentional barriers caused by the strain modulation and the amplitude of the shuttling pulse applied to the QuBus is too small, the position of the electron may become undetermined during shuttling, in particular for TiN devices under high stress. We note that electron shuttling in a QuBus device has been shown with a typical QD confinement energy in Ref. [10] but there Pt electrodes were used and some signature of tunneling were found during the conveyor-mode shuttle process. [12] Therefore a tailored shaping of the periodic input signals of the shuttle device may be necessary to compensate this effect.

To gain a better physical insight of the electrode-induced strain we numerically assessed the spatial distribution of the lattice deformation. To this aim we modeled the heterostructure and the TiN electrodes with the Solid Mechanics module of the *COMSOL Multiphysics* suite in order to simulate the electrode relaxation process through 2D simulations based on the Finite Element Method (FEM).

It is well established that both the intrinsic stress $\sigma$, [28] and the Young's Modulus $E$ of sputtered TiN depend on the deposition parameters. [50, 51] Since these quantities are required as input parameters for the mechanical modelling, we have preliminarily estimated the stress in the TiN electrodes from sample curvature measurements by XRD as $\sigma_A = -2.6 \pm 0.5$ GPa and $\sigma_B = -1.5 \pm 0.5$ GPa and their Young's Modulus by nanoindentation as $E_A = 106 \pm 25$ GPa and $E_B = 91 \pm 19$ GPa. As detailed in the SM, we have fed FEM simulation with this data to calculate the deviations induced in the multilayer stack with respect to the average biaxial strain, which results from the relaxation of the TiN gates.

Our main results are summarized in **Fig. 6** where we show in the yz plane the deviations of $\varepsilon_{yy}$ and $\varepsilon_{zz}$ from the average biaxial strain, calculated for a 1 µm-wide TiN fcontact pad (**a,b**) and for an array of clavier electrodes featuring a width of 160 nm and a pitch of 140 nm between them (**c,d**).

For the wide pad, we observe that the in-plane strain, shown in **panel A,** is more positive underneath the center of the TiN slab, representing an expansion of the lattice along the in-plane direction, which propagates down to a depth of > 300 nm below the sample surface. This behavior is in agreement with the positive in-plain strain modulation observed by SXDM in the Si QW layer under the wide (> 1 µm) TiN pads. The modulation of $\varepsilon_{zz}$ underneath the wide pad (**panel B**) is negative in the same region, and thus opposite in sign to $\varepsilon_{yy}$, as expected due to the Poisson effect and likewise observed

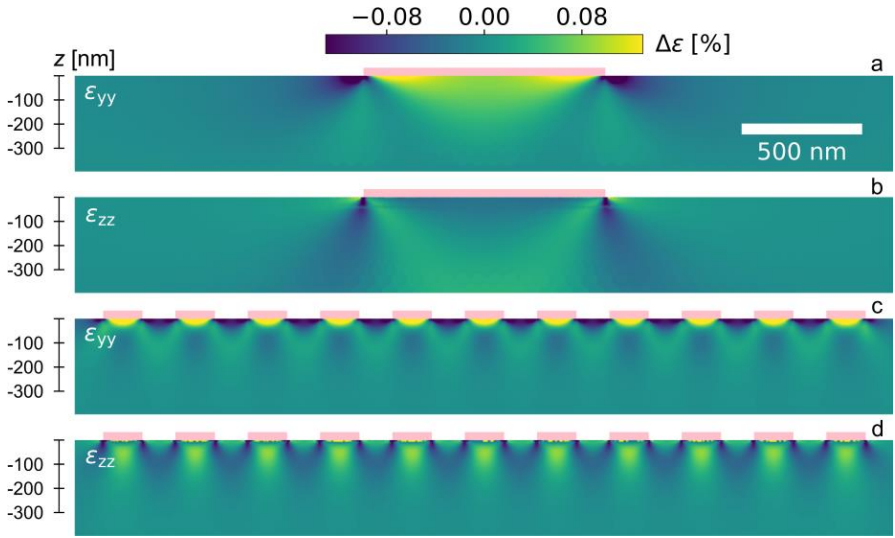

FIG. 6. Simulated maps in the vertical plane yz of the TiN-induced modulation of the diagonal strain components around the device structures: Maps around a TiN contact pad of 1 µm width of **(a)** $\varepsilon_{yy}$ and **(b)** $\varepsilon_{zz}$; maps around the QuBus clavier region in sample A featuring 160 nm wide electrodes with 140 nm pitch of **(c)** $\varepsilon_{yy}$ and **(d)** $\varepsilon_{zz}$. The TiN electrodes are symbolized by the magenta bars on the sample surface, the scale is the same for all images.

underneath wide pads in the SXDM strain maps.

Also below the narrow clavier electrodes, at a shallow depth of ~30 nm, the modulation of the in-plane strain $\varepsilon_{yy}$ induced in the SiGe material (**Fig. 6c**) is positive, due to the downward propagation of the tensile deformation experienced by the gates once they partially release their initial compressive stress. However, deeper in the sample we observe an inversion of the sign for $\varepsilon_{yy}$ which becomes negative, resulting in a compression of the lattice along the in-plane direction. In other words, and quite surprisingly, directly underneath a narrow compressively stressed electrode, we observe a compressive modulation of $\varepsilon_{yy}$. To qualitatively understand this counterintuitive effect, we observe that the tensile field from the bottom surface of the electrode propagate downward mainly from its edges and at an angle of ~45°, in line with numerical results reported in Ref. [52] for compressively strained SiGe stripes deposited on Si. On the other hand, the tensile field in the central electrode region does not propagate effectively downward. Consequently, starting from a certain depth the central material region results to be sandwiched between two tensile strained domains which induce its compressive distortion.

Having in mind this kind of phenomenology, we can better understand the SXDM strain maps, where we observe that underneath the clavier electrodes, the modulations in $\varepsilon_{yy}$ are negative (compressive) rather than positive (tensile). We therefore conclude that this depth-dependent strain inversion obtained by FEM may be the reason for the qualitative difference between the wide and the narrow structures in the experimental data. A similar inversion effect is observed also for the $\varepsilon_{zz}$ strain component underneath the narrow electrodes, as shown in **panel D**. Indeed, at shallow depth, it is negative underneath each clavier electrode, but becomes positive at larger depth, in line with the expected opposite behavior of $\varepsilon_{yy}$ and $\varepsilon_{zz}$.

## IV. DISCUSSION

In this work, we investigated the effect of CMOS-processed TiN electrodes on the lateral stain distribution in a 10 nm thin Si QW. For several components of the strain tensor, local modulations of $2 - 8 \times 10^{-4}$ due to the stressing action of the TiN stripes were observed. By comparison of the global stress determined from the sample bow in extended TiN films to the strain modulations caused by the small electrodes, we find that the macroscopic stress in the gate material is proportional to the microscopic strain induced by the electrodes. By bandstructure calculations, we translated the electrode-driven strain modulations into local changes of the potential level of the conduction band minimum with a bandwidth of several meV, similar to the orbital energy of an electrostatic QD hosting a few electrons. [10] These may act as unintentional barriers hindering spin-coherent shuttling of electrons in conveyer mode, [22] as well as affect charging energies and gate voltages required to operate spin qubits in large arrays. [24, 23]

We furthermore observe size effects of the electrode-induced strain modulations in the QW layer. By 2D FEM simulations we find that underneath a thin compressively stressed electrode, an inversion of the sign of the strain occurs 50-70 nm to the sample surface. This supports the experimental strain maps determined by SXDM, where we observe strain modulations of different signs underneath TiN stripes of 160 nm and >1 µm lateral extent, respectively. We predict that, by placing the QW layer at the exact depth of the inversion point for a given device geometry, devices may be fabricated in which the electrode-driven strain modulations are strongly reduced.

It is worth noting that, when cooled down to typical operation temperatures in a quantum processor of approx. 20 mK, the strain landscape changes due to the different coefficients of thermal expansion of all the materials in the sample. [53, 54] This additional thermal strain needs to be taken into account when employing strain-engineering to improve the device. Elastic effects may be predicted by a suitable FEM model, which needs to be benchmarked against an experimental reference point for the initial stress condition. Moreover, also inelastic relaxation processes could take place during thermal cycling.

We expect that a more detailed characterization will be required in the future on functional QuBus devices. Moreover, we anticipate that similar modulations of the spatial landscapes of strain and potential occur in other technology platforms for quantum computing featuring micron-scale lithographic structures, such as superconducting qubits, [55] as well as different types of microelectronic devices. [2] The impact of electrode-driven strain may be even more significant in Metal-Oxide-Semiconductor (MOS) type devices, in which the quantum dots are located directly underneath the interface between the electrode and the oxide. In particular shear strain, which is difficult to measure with other experimental techniques, has been shown to affect the nuclear electric resonance and the quadrupole interaction of a coherently controlled nuclear spin in quantum MOS devices. [56]

In summary, we demonstrate that the strain induced in a thin Si QW layer by gate electrodes depends not only on the stress in the gate material, but also the lateral dimensions of the electrodes and the depth of the QW layer in the sample. We predict that this effect is of a magnitude sufficient foto affect the performance of quantum devices, such as coherent electron shuttles, but may be avoided by placing the QW layer at a depth corresponding to a "geometrical sweet spot". Moreover, our results highlight the potential of SXDM as a non-destructive technique to study complex strain distributions in microscopic devices and to access important material related information for microelectronics and quantum technology.

## Supplemental Material

See the Supplemental Material for a description of the SXDM data analysis, Maps of lattice strain and composition in the SiGe buffer layer, additional results from Raman-microspectroscopy strain mapping, nanoindentation measurements, film stress measurement by sample curvature, details of FEM simulations and bandstructure calculations, ptychographic reconstruction of the X-ray nanoprobe.

## ACKNOWLEDGMENTS

We acknowledge the European Synchrotron Radiation Facility for provision of synchrotron radiation facilities and the staff for assistance in using beamline ID01.

Furthermore, the authors thank Dr. J. Martin from IKZ for fabricating fluorescent markers, Dr. T. U. Schulli from ESRF for setting up the fluorescence detector, Dr. Y. Liu from IKZ for proof reading the manuscript, P. Muster from Infineon Technologies Dresden GmbH for useful discussion, Dr. I. Herman from Eurofins EAG GmbH for providing nanoindentation measurements, and the cleanroom staffs at IHP and at Infineon Dresden GmbH for technical support and processing wafers.

This research was supported by the Leibniz-associations "High-definition crystalline Silicon-Germanium structures for Quantum Circuits" project (SiGeQuant, project number K124/2018), the European Union and its Horizon 2020 framework program (FETFLAG-05-2020) as part of the project "Quantum Large Scale Integration in Silicon" (QLSI, grant agreement number 951852 ) and the German Federal Ministry of Education and Research within the frame of the project "Halbleiter-Quantenprozessor mit shuttlingbasierter skalierbarer Architektur" (QUASAR, FKZ: 13N15654).

## AUTHOR DECLARATIONS

The authors have no conflicts of interest to disclose.

## Author Contributions

**Cedric Corley-Wiciak:** Investigation (equal), Formal Analysis (equal), Writing/Original Draft Preparation (lead), Visualization (lead), **M. H. Zoellner:** Conceptualization (equal), Investigation (equal), Visualization (supporting), Formal Analysis (supporting) **I. Zaitsev:** Methodology (supporting), Validation (equal) Writing/Original Draft Preparation (supporting) **K. Anand:** Resources (equal), **E. Zatterin:** Investigation (equal), Software (support), Methodology (equal), **Y. Yamamoto:** Resources (equal), supervision (supporting), **A. A. Corley-Wiciak:** Investigation (supporting), Visualization (supporting) **F. Reichmann:** Funding Acquisition (equal), Writing/Review and Editing (supporting), **W. Langheinrich:** Resources (equal), Writing/Review and Editing (supporting) **L. R. Schreiber:** Funding Acquisition (equal), Writing/Review and Editing (equal), Conceptualization (equal), **C. L. Manganelli:** Software (equal), Supervision (equal), Validation (equal), **M. Virgilio:** Validation (lead), Software (equal), Writing/Review and Editing (equal), **C. Richter:** Investigation (lead), Data Curation (lead), Software (equal), Methodology (lead) **G. Capellini:** Funding Acquisition (lead), Writing/Review and Editing (lead), Supervision (lead), Conceptualization (lead)

## DATA AVAILABILITY STATEMENT